\documentstyle[aps,prl,psfig,floats]{revtex}

\begin{document}
\draft
\wideabs{

\title{Edge states and determination of pairing symmetry in
       superconducting $\bf Sr_2RuO_4$} 

\author{K. Sengupta, Hyok-Jon Kwon, and Victor M. Yakovenko}

\address{Department of Physics and Center for Superconductivity
   Research, University of Maryland, College Park, MD 20742-4111}

\date{\bf cond-mat/0106198, v.1: 11 June 2001; v.2: 2 September; v.3:
20 November; v.4: 11 January 2002}


\maketitle

\begin{abstract}
We calculate the energy dispersion of the surface Andreev states and
their contribution to tunneling conductance for the order parameters
with horizontal and vertical lines of nodes proposed for
superconducting $\rm Sr_2RuO_4$.  For vertical lines, we find double
peaks in tunneling spectra reflecting the van Hove singularities in
the density of surface states originating from the turning points in
their energy dispersion.  For horizontal lines, we find a single
cusp-like peak at zero bias, which agrees very well with the
experimental data on tunneling in $\rm Sr_2RuO_4$.
\end{abstract}

 }

\section{Introduction}

Numerous experiments suggest that the superconducting state of $\rm
Sr_2RuO_4$ \cite{Maeno94} is unconventional (see review
\cite{PhysToday}).  Strong suppression of $T_c$ by non-magnetic
impurities \cite{impurities} and absence of the Hebel-Slichter peak in
NMR experiments \cite{Ishida97} indicate that it is not $s$-wave.  The
absence of spin susceptibility reduction below $T_c$ in the Knight
shift measurements indicates the spin-triplet pairing \cite{Ishida98}.
The $\mu$SR experiment suggests a superconducting state with broken
time-reversal symmetry \cite{Luke98}.  Small-angle neutron scattering
reveals a square vortex lattice, which is interpreted as the
indication for a two-component order parameter \cite{Riseman98}.
These experiments initially led to suggestion of the 2D chiral
(time-reversal non-invariant), isotropic, nodeless $p$-wave pairing
potential \cite{Rice95}.  However, the power-law temperature
dependences found in specific heat \cite{Nishi00}, nuclear relaxation
rate \cite{Ishida00}, penetration depth \cite{Bonalde00}, thermal
conductivity \cite{Tanatar00}, and ultrasonic attenuation
\cite{Matsui01,Lupien01} indicate nodes in the energy gap.  In
response to these experiments, the following alternative order
parameters were proposed \cite{Radtke}: anisotropic $p$-wave
\cite{Miyake99}, $p$-wave with horizontal lines of nodes
\cite{Hasegawa00}, and $f$-waves with vertical
\cite{Hasegawa00,Balatsky} or horizontal lines of nodes \cite{Maki-W}.
It was also proposed that the $\alpha$ and $\beta$ bands of $\rm
Sr_2RuO_4$ are either not superconducting \cite{Agterberg97} or have
horizontal lines of nodes \cite{Zhitomirsky}.  Big in-plane anisotropy
of ultrasound attenuation \cite{Lupien01} may support the vertical
lines of nodes.  However, thermal conductivity depends very little on
the orientation of an in-plane magnetic field \cite{Izawa00}, which is
against the vertical lines \cite{Maki-DW}.

Electron tunneling between a normal metal and a superconductor proved
to be an important tool in determining superconducting symmetry.
Observation of the zero-bias conductance peak (ZBCP) \cite{ZBCP} due
to the formation of the midgap Andreev bound states \cite{Hu94}
confirmed the $d$-wave symmetry of the high-$T_c$ cuprates.  Electron
tunneling into a three-dimensional $p$-wave superconductor with the
pairing potential corresponding to the B-phase of $^3$He was
considered in Ref.\ \cite{Buchholtz81}.  Electron tunneling in $\rm
Sr_2RuO_4$ was studied theoretically in Refs.\ \cite{Tanaka97} and
\cite{Honerkamp98} for the 2D isotropic chiral unitary and non-unitary
$p$-waves.  In this paper, we calculate the tunneling conductance
curves for the alternative order parameters listed above and compare
them with experiments \cite{Laube00,Mao01}.

\section{General formalism}

\begin{figure}
\centerline{\psfig{file=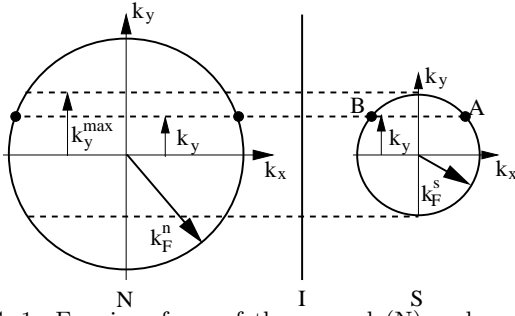,width=0.8\linewidth,angle=0}}
\caption{Fermi surfaces of the normal (N) and superconducting (S)
  materials forming the interface I.  The points A and B on the Fermi
  surface S are connected by specular reflection from the interface.}
\label{fig:Fermi}
\end{figure}

We model the tunneling contact by two semi-infinite regions, normal
(N) and superconducting (S), with a flat interface (I) perpendicular
to the $a$ axis of $\rm Sr_2RuO_4$ (see Fig.\ \ref{fig:Fermi}).  The
$x$ and $y$ axes are selected along the crystal axes $a$ and $b$,
respectively.  The tunneling conductance is calculated at zero
temperature by solving the Bogoliubov-de Gennes equations in the
ballistic regime following Refs.\ \cite{Blonder82,Tanaka95}.  The
tunneling barrier is modeled by a delta-function potential of strength
${\cal H}$, so the boundary conditions at the interface are
$\psi_n|_I=\psi_s|_I$ and
$\hat{v}_n\psi_n|_I=\hat{v}_s\psi_s|_I-2i{\cal H}\psi_n|_I$.  Here
$\psi_n$ and $\psi_s$ are the electron wave functions in the normal
and superconducting regions, and $\hat{v}_{n,s}$ are the velocity
components perpendicular to the interface.  The pairing potential of a
triplet superconductor can be expressed as
$\hat\Delta=i\hat\sigma_y(\hat{\bbox\sigma}\cdot{\bf d})\Delta({\bf
k},x)$, where $\hat{\bbox\sigma}$ are the Pauli matrices operating on
the electron spin indices, the vector ${\bf d}$ indicates the
direction of spin polarization, and ${\bf k}$ is the relative momentum
of electrons in a Cooper pair.  In this paper, we consider the cases
where ${\bf d}$ does not depend on ${\bf k}$ and is pinned to the $c$
axis of $\rm Sr_2RuO_4$. The spatial dependence of the pairing
potential is taken to be step-like: $\Delta({\bf k},x)=\Delta({\bf
k})\,\Theta(x)$.  Assuming that the electron momentum $k_y$ parallel
to the interface is conserved, the tunneling conductance per one
layer, $\overline{G}(V)$, can be written as an integral over $k_y$
\cite{Blonder82,Tanaka95}:
\begin{eqnarray}
  \overline{G}(V) &=& \frac{2e^2}{h}\,l
  \int_{-k_y^{\rm max}}^{k_y^{\rm max}}\frac{dk_y}{h}\,G(V,k_y),
\label{tcond}\\ 
  G(V,k_y) &=& D\,\frac{1+|\Gamma|^2\,D-(1-D)\,|\Gamma|^4} {\Big|
  1-(1-D)\,\Gamma^2\,\exp(i\Phi) \Big|^2}.
\label{tunnel1}
\end{eqnarray}
In Eq.\ (\ref{tcond}), $e$ is the electron charge, $h$ is the Planck
constant, $l$ is the length of the interface, and $V$ is the bias
voltage.  We consider the case where the Fermi surfaces of both the
normal metal and superconductor are circular, and their radii are
$k_F^{(n)}>k_F^{(s)}$, as shown in Fig.\ \ref{fig:Fermi}.  The limits
of integration are set by the smaller Fermi momentum: $k_y^{\rm
max}=k_F^{(s)}$.  In Eq.\ (\ref{tunnel1}),
$D(k_y)=4v_n(k_y)v_s(k_y)/\{[v_n(k_y)+v_s(k_y)]^2+4{\cal H}^2\}$ is
the normal-state transmission coefficient, where $v_{n,s}(k_y)$ are
the Fermi velocities components perpendicular to the interface.
$\Phi(k_y)=\phi_A(k_y)-\phi_B(k_y)$ is the phase difference of the
superconducting pairing potentials at the points on the Fermi surface
connected by electron reflection from the interface (points A and B in
Fig.\ \ref{fig:Fermi}) \cite{Tanaka95}.  $\Gamma(V,k_y)$ is
\begin{eqnarray}
 \Gamma &=& \frac{eV-{\rm sgn}(eV)\sqrt{(eV)^2-|\Delta(k_y)|^2}}
   {|\Delta(k_y)|}, \quad |eV|\geq|\Delta|, \nonumber \\
 \Gamma &=& \frac{eV-i\sqrt{|\Delta(k_y)|^2-(eV)^2}}{|\Delta(k_y)|},
   \quad |eV|\leq|\Delta|.
\label{factor}
\end{eqnarray}

Denoting $G(V,k_y)$ as ${\cal G}(V,k_y)$ for the subgap voltage
$|eV|\leq|\Delta|$, where $|\Gamma(V)|=1$, we rewrite Eq.\
(\ref{tunnel1}) as
\begin{eqnarray}
&& {\cal G}(V,k_y) = \frac{D^2(k_y)/2[1-D(k_y)]}
  {D^2(k_y)/4[1-D(k_y)] + F^2(V,k_y)},
  \label{tunnel2} \\
&& F=\sqrt{1-|eV/\Delta|^2}\cos(\Phi/2)-(eV/|\Delta|)\sin(\Phi/2).
\label{F}
\end{eqnarray}
For a high barrier, $D\approx v_nv_s/{\cal H}^2\ll1$.  Then Eq.\
(\ref{tunnel2}) becomes ${\cal G}(V,k_y)\approx\pi
D(k_y)\,\delta[F(V,k_y)]$, and Eq.\ (\ref{tcond}) gives
\begin{equation}
  \overline{\cal G}(V) \approx C
  \int^{k_y^{\rm max}}_{-k_y^{\rm max}}
  dk_y\,v_n(k_y)\,v_s(k_y)\,\delta[F(V,k_y)],
\label{tuncond0}
\end{equation}
where $C=2\pi e^2l/h^2{\cal H}^2$.  The delta-function contributes to
the integral (\ref{tuncond0}) when $F(V,k_y)=0$.  Using Eq.\
(\ref{F}), this condition can be written as $eV\!=\!E(k_y)$, where
\begin{equation}
  E(k_y) = |\Delta(k_y)|\cos[\Phi(k_y)/2]
\label{energy} 
\end{equation}
for $\sin(\Phi/2)\ge0$, i.e.\ $0\le\Phi\le2\pi$, which is the
appropriate interval for $\Phi$.  $E(k_y)$ (\ref{energy}) is nothing
but the energy of a surface Andreev state with the momentum $k_y$
obtained for the impenetrable barrier (${\cal H}\rightarrow\infty$)
\cite{Hu94}.  The energy density of these surface states (DOS) is
\begin{equation} 
  \rho(\epsilon) = \int\limits_{ -k_F^{(s)} }^{ k_F^{(s)} }
  \frac{dk_y}{h} \, \delta[\epsilon-E(k_y)]
  =\sum_j\frac{1}{h\,|\partial_{ k_y^{(j)} }E( k_y^{(j)} )|},
\label{doseq}
\end{equation}
where $k_y^{(j)}$ is the $j$-th root of the equation $E(k_y)=\epsilon$.

The delta-function in Eq.\ (\ref{tuncond0}) indicates that the subgap
tunneling takes place when the bias voltage matches the energy of a
surface state: $eV=E(k_y)$.  Denoting the $j$-th root of this equation
as $k_y^{(j)}$ and resolving the delta-function in Eq.\ (\ref{tuncond0}),
we find \FL
\begin{equation}
  \overline{\cal G}(V) \approx C \sum_j
  \frac{v_n(k_y^{(j)}) v_s(k_y^{(j)}) \sqrt{|\Delta(k_y^{(j)})|^2-(eV)^2}}
  {|\partial_{ k_y^{(j)} } E( k_y^{(j)} ) |}.
\label{tuncond}
\end{equation}
Because the denominators in Eqs.\ (\ref{tuncond}) and (\ref{doseq})
are the same, both DOS and tunneling conductance exhibit peaks (the
inverse-square-root van Hove singularities) at the turning points of
the energy dispersion, where $\partial_{k_y}E(k_y)=0$.  The peak
positions depend solely on the pairing potential $\Delta(k_y)$ through
Eq.\ (\ref{energy}) and not on the band structure details.  The number
of such turning points does not change upon small continuous
deformation of $\Delta(k_y)$, so it is a topological feature.
However, because of the additional factors in Eq.\ (\ref{tuncond}),
tunneling conductance is not simply proportional to DOS, as often
assumed.  Eq.\ (\ref{tuncond}) vanishes at $k_y\to k_F^{(s)}$, where
the normal component $v_s$ of the Fermi velocity goes to zero.  It
also vanishes when $\Delta(k_y)\to0$ or $E(k_y)\to\pm\Delta(k_y)$
(i.e.\ $\Phi(k_y)\to0,\: 2\pi$).  In these cases, the quasiparticle
localization length perpendicular to the interface, $\lambda=\hbar
v_s/\sqrt{|\Delta|^2-E^2}$, diverges \cite{Hu94}; thus, the
probability to find a quasiparticle on the surface vanishes.

\section{Application to $\bf S\lowercase{r}_2R\lowercase{u}O_4$ }

\begin{figure}
\centerline{\psfig{file=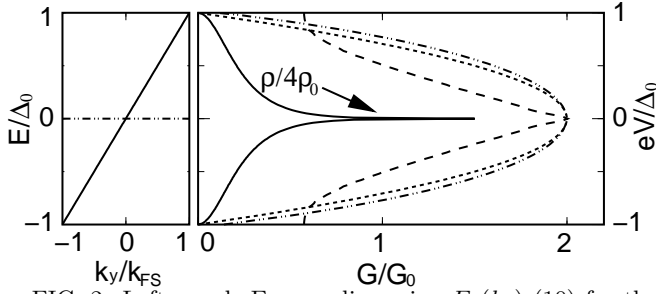,width=\linewidth,angle=0}}
\caption{Left panel: Energy dispersion $E_1(k_y)$ (\ref{D1}) for the
  2D isotropic $p$-wave. Right panel: The corresponding exact [Eqs.\
  (\ref{tcond}) and (\ref{tunnel1}), dash-dotted line] and approximate
  [Eq.\ (\ref{G0}), dotted line] tunneling conductance curves; DOS
  (solid line) and tunneling conductance (long-dashed line) for the
  $p$-wave with horizontal lines of nodes [Eq.\ (\ref{D5})].}
\label{p1fig}
\end{figure}

\begin{figure}
\centerline{\psfig{file=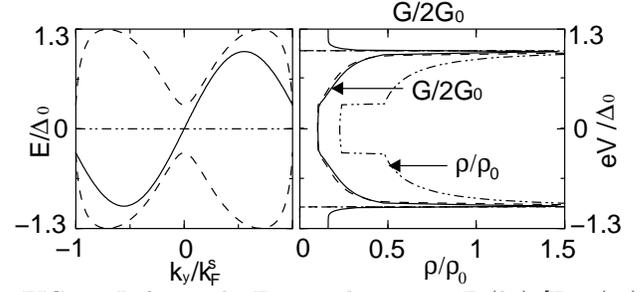,width=0.95\linewidth,angle=0}}
\caption{Left panel: Energy dispersion $E_2(k_y)$ [Eq.\ (\ref{E2}),
  solid line] and gap $|\Delta_2(k_y)|$ [Eq.\ (\ref{D2}), dashed line]
  for the anisotropic $p$-wave.  Right panel: the corresponding DOS
  [dashed-dotted line] and the exact (solid line) and approximate
  [dashed line] tunneling conductance curves.}
\label{p2fig} 
\end{figure}

In Figs.\ \ref{p1fig}, \ref{p2fig}, \ref{f1fig}, and \ref{f2fig}, we
show how this general formalism applies to the alternative
superconducting order parameters proposed for $\rm Sr_2RuO_4$ in the
literature.  In order to focus on the features of superconductor,
rather than normal metal, we took $k_F^{(n)}=5k_F^{(s)}$.  Then, the
perpendicular velocity of the normal metal is approximately constant
for all momenta $|k_y|\le k_F^{(s)}$: $v_n\approx v_F^{(n)}$, while in
the superconductor it is $v_s=v_F^{(s)}\sqrt{1-(k_y/k_F^{(s)})^2}$,
where $v_F^{(n)}$ and $v_F^{(s)}$ are the corresponding Fermi
velocities.  The numerical calculations are done for the high barrier
$Z={\cal H}/\sqrt{ v_F^{(n)}v_F^{(s)} }=3$.  As the figures
illustrate, in this limit, the approximate curves given by Eq.\
(\ref{tuncond}), which includes only the contribution of the localized
surface states, agree well with the exact curves obtained from Eqs.\
(\ref{tcond}) and (\ref{tunnel1}), which also include the contribution
of the extended bulk states.

\subsection{Nodeless pairing potential }

For the 2D isotropic chiral $p$-wave, the pairing potential and the
energy dispersion of the edge states are \cite{Rice95,Honerkamp98}
\begin{equation}
  \Delta_1({\bf k})=\Delta_0(k_x +ik_y)/k_F^{(s)}, 
  \quad E_1(k_y)=\Delta_0k_y/k_F^{(s)}.
\label{D1} 
\end{equation}
Substituting Eq.\ (\ref{D1}) into Eqs.\ (\ref{doseq}) and
(\ref{tuncond}), we find
\begin{eqnarray}
  &&\rho(\epsilon)=k_F^{(s)}/h\Delta_0\equiv\rho_0,
\label{rho0} \\
  &&\overline{\cal G}(V) \approx 2G_0
  [1-(eV/\Delta_0)^2],
\label{G0}
\end{eqnarray}
where $G_0=\pi e^2lk_F^{(s)}v_F^{(n)}v_F^{(s)}/h^2{\cal{H}}^2$ is the
normal-state conductance.  The energy dispersion $E_1(k_y)$ (\ref{D1})
has no turning points, and the corresponding DOS (\ref{rho0}) is flat:
$\rho(\epsilon)=\rm const$.  The tunneling conductance curve
$\overline{G}(V)$ (\ref{G0}) is parabolic for $|eV|<\Delta_0$,
attaining the maximal value $2G_0$ at $V=0$ and vanishing at
$|eV|=\Delta_0$, as shown in Fig.\ \ref{p1fig}.  For $|eV|>\Delta_0$,
$\overline{G}(V)$ increases with increasing $|V|$ and saturates at the
normal-state value $G_0$ at $|eV|\gg\Delta_0$.  This behavior is in
agreement with previous numerical calculations
\cite{Tanaka97,Honerkamp98}.  One can recognize that the curves shown
in Fig.\ 4 of Ref.\ \cite{Honerkamp98} and the corresponding figures
of Ref.\ \cite{Tanaka97} for $|eV|<\Delta_0$ are actually distorted
parabolas.  The deviations are due to the finite values of $Z$ and to
the finite size of the acceptance cone utilized in the numerical
calculations \cite{Tanaka97,Honerkamp98}.  Our analytical expression
(\ref{G0}) applies in the case $Z\gg1$ and for the full acceptance
cone.

\subsection{Pairing potentials with vertical lines of nodes }

In this subsection, we consider pairing potentials with vertical lines
of nodes, which originate from modulation of the gap function
$\Delta({\bf k})$ in the $(k_x,k_y)$ plane.

For the anisotropic chiral $p$-wave proposed in Ref.\ \cite{Miyake99},
the pairing potential and the energy dispersion are
\begin{eqnarray}
  \Delta_2({\bf k}) &=& \Delta_0[\sin(k_xa/\hbar)+i\sin(k_y a/\hbar)], 
\label{D2} \\
  E_2(k_y) &=& \Delta_0\sin(k_ya/\hbar),
\label{E2}
\end{eqnarray}
where $a$ is the lattice constant, and $k_F^{(s)}a=0.9\pi\hbar$.
Substituting Eqs.\ (\ref{D2}) and (\ref{E2}) into Eqs.\ (\ref{tcond}),
(\ref{tunnel1}), (\ref{factor}), (\ref{tuncond}) and (\ref{doseq}), we
calculate the tunneling conductance and DOS curves shown in Fig.\
\ref{p2fig}.  They exhibit peaks at $eV=\pm\Delta_0$ originating from
the turning points in the edge states dispersion.  Equation
$eV=E(k_y)$ has one solution $k_y^{(j)}$ for $|eV|<0.31\,\Delta_0$ and
two solutions for $0.31\,\Delta_0<|eV|<\Delta_0$.  The switch causes
the discontinuity of DOS and the slope change of tunneling conductance
at $eV=\pm0.31\,\Delta_0$.

\begin{figure}[b]
\centerline{\psfig{file=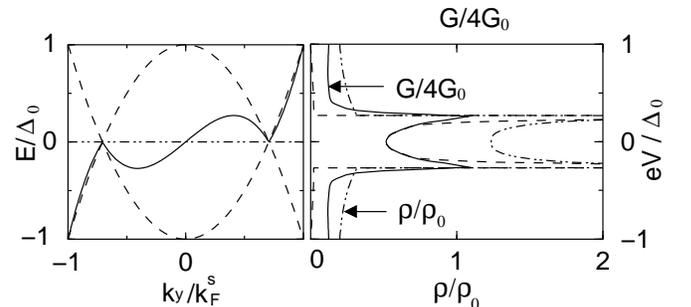,width=\linewidth,angle=0}}
\caption{The same curves as in Fig.\ \ref{p2fig} for the $f$-wave
  pairing potential (\ref{D3}) with vertical lines of nodes.}
\label{f1fig}
\end{figure}

For the $f$-wave proposed in Ref.\ \cite{Hasegawa00}, we have
\begin{eqnarray}
  \Delta_3({\bf k})&=&\Delta_0\,(k_x^2-k_y^2)\,(k_x+ik_y)\,/\,(k_F^{(s)})^3,
\label{D3} \\
  E_3(k_y)&=&\Delta_0\,|k_x^2-k_y^2|\,k_y\,/\,(k_F^{(s)})^3.
\label{E3}
\end{eqnarray}
As shown in Fig.\ \ref{f1fig}, the edge states dispersion has cusps at
$k_y=\pm k_F^{(s)}/\sqrt{2}$, where $\Delta_3({\bf k})$ vanishes, and
two turning points in between.  Consequently, the tunneling
conductance and DOS curves have peaks at $eV=\pm0.26\,\Delta_0$.

\begin{figure}
\centerline{\psfig{file=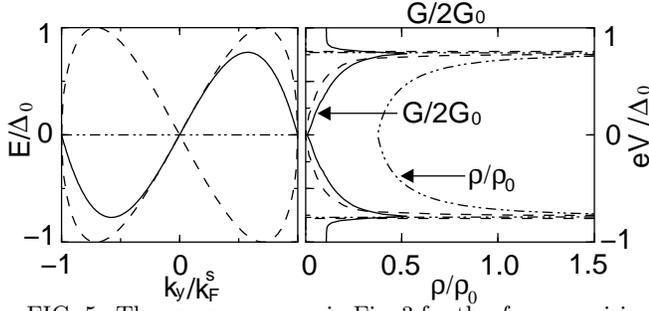,width=\linewidth,angle=0}}
\caption{The same curves as in Fig.\ \ref{p2fig} for the $f$-wave
  pairing potential (\ref{D4}) with vertical lines of nodes.}
\label{f2fig}
\end{figure}

For another $f$-wave proposed in Refs.\ \cite{Hasegawa00,Balatsky}, we
have
\begin{eqnarray}
  \Delta_4({\bf k})&=&2\Delta_0\,k_xk_y\,(k_x+ik_y)\,/\,(k_F^{(s)})^3,
\label{D4} \\
  E_4(k_y)&=&2\Delta_0\,k_x^{2}\,k_y\,/\,(k_F^{(s)})^3.
\label{E4}
\end{eqnarray}
Eqs.\ (\ref{D4}) and (\ref{E4}) transform into Eqs.\ (\ref{D3}) and
(\ref{E3}) upon a $\pi/4$ rotation in the $x$-$y$ plane.  As shown in
Fig.\ \ref{f2fig}, the turning points in the energy dispersion and the
peaks in the tunneling conductance and DOS curves occur at
$eV=\pm0.76\,\Delta_0$.  Their position is approximately the same as
in Fig.\ \ref{p2fig} for the anisotropic $p$-wave.  However, in Fig.\
\ref{f2fig}, unlike in all other figures, tunneling conductance
vanishes at zero bias, even though DOS remains finite.  This
characteristic feature of the $f$-wave potential (\ref{D4}) is a
consequence of the square-root factor in Eq.\ (\ref{tuncond}).

Thus, tunneling conductance $\overline{G}(V)$ exhibits two peaks
located symmetrically around $V=0$ for all pairing potentials with
vertical lines of nodes described in this subsection.  This conclusion
is in agreement with recent numerical calculations \cite{Stefanakis}.

\subsection{Pairing potentials with horizontal lines of nodes }

Thus far we considered pairing potentials that do not depend on $k_z$,
the momentum component perpendicular to the $\rm Sr_2RuO_4$ planes.
Ref.\ \cite{Hasegawa00} also proposed the $p$-wave with horizontal
lines of nodes at certain $k_z$.  Its pairing potential can be
obtained from Eq.\ (\ref{D1}) by the following substitution
\begin{equation}
  \Delta_0\to\tilde\Delta_0(p_z)=\Delta_0\cos(p_z),
\label{D5}
\end{equation}
where we switched to the dimensionless variable $p_z=k_zc/\hbar$ ($c$
is the interplane distance).

Now DOS is obtained by averaging $\rho(\epsilon)$ from Eq.\
(\ref{rho0}) over $p_z$ using Eq.\ (\ref{D5}):
\begin{eqnarray}
  \rho(\epsilon)&=&\int\limits_{|\tilde\Delta_0(p_z)|>|\epsilon|}
  \frac{dp_z}{2\pi} \frac{k_F}{h\,|\tilde\Delta_0(p_z)|}
\label{rho5} \\
  &=& \frac{\rho_0}{\pi}\ln\frac{1+\sqrt{1-(\epsilon/\Delta_0)^2}}
  {1-\sqrt{1-(\epsilon/\Delta_0)^2}}.
\label{rho5exact}
\end{eqnarray}
In Eq.\ (\ref{rho5}), $\tilde\Delta_0(p_z)$ vanishes linearly with
$p_z$ at the nodes, which results in a logarithmic divergence of
$\rho(\epsilon)$ at $\epsilon\to0$:
\begin{equation}
\rho(\epsilon)
\propto\ln|\Delta_0/\epsilon|, \quad {\rm for} 
\quad |\epsilon|\ll\Delta_0.
\label{rho5approx}
\end{equation}
The logarithmic divergence is cut off at
$|\tilde\Delta_0(p_z)|=|\epsilon|$, because Eq.\ (\ref{rho0}) applies
only to the subgap states with $|\epsilon|<|\tilde\Delta_0(p_z)|$.
The plot of $\rho(\epsilon)$ is shown in Fig.\ \ref{p1fig}.

Tunneling conductance $\overline{\cal G}(V)$ is obtained by
substituting Eq.\ (\ref{D5}) into Eq.\ (\ref{G0}) and averaging over
$p_z$:
\begin{eqnarray}
  \overline{\cal G}(V)&=&2\,G_0\int\limits_{|\tilde\Delta_0(p_z)|>|eV|}
  \frac{dp_z}{2\pi}\left(1-
  \left|\frac{eV}{\tilde\Delta_0(p_z)}\right|^2\right)
\label{G5} \\
  &=&\frac{4}{\pi}\,G_0\left(
  \arccos\left|\frac{eV}{\Delta_0}\right|
  -\left|\frac{eV}{\Delta_0}\right|
  \sqrt{1-\left|\frac{eV}{\Delta_0}\right|^2}
  \right).
\label{G5exact}
\end{eqnarray}
In Eq.\ (\ref{G5}), the divergence of $1/|\tilde\Delta_0(p_z)|^2$ at
the node is cut off at $|eV|$, because Eq.\ (\ref{G0}) applies only
for $|eV|<|\tilde\Delta_0(p_z)|$.  To treat the divergence, it is
convenient to change the variable of integration from $p_z$ to
$\Delta$:
\begin{equation}
  \overline{\cal G}(V) = 2G_0\left(1
  - \frac{4}{2\pi}\int_0^{|eV|}\frac{d\Delta}{\Delta'}
  - \frac{4\,|eV|^2}{2\pi}\int_{|eV|}^{\Delta_0}\frac{d\Delta}{\Delta'\Delta^2}
  \right).
\label{G5'}
\end{equation}
Here the factors of 4 come from the two nodal lines and the
integration over the two sides of each line.  The slope
$\Delta'=|d\tilde\Delta_0(p_z)/dp_z|$ is approximately constant at the
nodes: $\Delta'\approx\Delta_0$.  Then the integrals in Eq.\
(\ref{G5'}) can be taken, and we find
\begin{equation}
  \overline{\cal G}(V) \approx 2G_0\left(1
  -\frac{4}{\pi}\left|\frac{eV}{\Delta_0}\right|
  \right) \quad {\rm for} \quad |eV|\ll\Delta_0.
\label{G5approx}
\end{equation}
Thus, tunneling conductance has a cusp at zero bias.  The curve for
$\overline{G}(V)$ calculated exactly starting from Eqs.\ (\ref{tcond})
and (\ref{tunnel1}) is shown in Fig.\ \ref{p1fig} and indeed
demonstrates the cusp in agreement with Eq.\ (\ref{G5approx}).

Notice the difference in Fig.\ \ref{p1fig} between the singular
logarithmical divergence of DOS and the triangular-shaped cusp in
tunneling conductance at zero bias.  Both features result from the
pile-up of surface Andreev states at zero energy caused by vanishing
gap at the nodes.  These are robust features of the chiral pairing
potentials with horizontal lines of nodes, independent of the band
structure details.  While Eqs.\ (\ref{rho5exact}) and (\ref{G5exact})
are derived specifically for the gap (\ref{D5}) being a cosine
function, the asymptotic Eqs.\ (\ref{rho5approx}) and (\ref{G5approx})
are valid for any generic chiral pairing potential with horizontal
lines of nodes.  For example, the $f$-wave with horizontal lines of
nodes \cite{Maki-W},
\begin{eqnarray}
  \Delta_6({\bf k})&=&i\Delta_0\,\sin(p_z)\,
  (k_x+ik_y)^2\,/\,(k_F^{(s)})^2,
\label{D6}\\
  E_6(k_y)&=&{\rm sgn}(k_y)\,\Delta_0\,[ 2(k_y/k_F^{(s)})^2 -1]\,
  |\sin(p_z)|,
\label{E6}
\end{eqnarray}
produces similar curves for the in-plane tunneling.  However, being an
odd function of $p_z$, it also exhibits a ZBCP in the $c$-axis
tunneling.

\section{Comparison with experiment}

Point-contact tunneling spectroscopy \cite{Laube00} found a ZBCP in
$\rm Sr_2RuO_4$.  The data were fitted using the 2D isotropic $p$-wave
(\ref{D1}) and assuming a narrow acceptance cone, i.e.\ a small value
of $k_y^{\rm max}$.  In this situation, only the states with
$k_y\approx0$ and hence $E(k_y)\approx0$ are probed; thus all pairing
potentials, except the $f$-wave (\ref{D4}), would show a ZBCP.
However, such a ZBCP, in contrast to nonchiral cuprates \cite{Hu94}
and organic superconductors \cite{Sengupta00}, does not identify the
pairing symmetry uniquely.

\begin{figure}
\centerline{\psfig{file=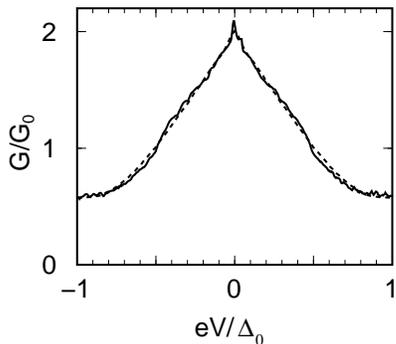,width=0.6\linewidth,angle=0}}
\caption{Long-dashed line: Theoretical tunneling conductance curve
  from our Fig.\ \ref{p1fig} calculated for the $p$-wave with
  horizontal lines of nodes [Eq.\ (\ref{D5})] using $\Delta_0=0.8$
  meV.  Solid line: Experimental curve from Fig.\ 2 of Ref.
  \protect\cite{Mao01} for 0.8 K.}
\label{fig:exper}
\end{figure}

The more recent tunneling experiments \cite{Mao01} were performed on
cleaved junctions of $\rm Sr_2RuO_4$ with inclusions of Ru.  Although
nominally their setup corresponds to tunneling along the $c$ axis, the
experimentalists believe that the actual tunneling takes place in the
$(a,b)$ plane via the inclusions of Ru.  They found a cusp-like peak
at zero bias in the $1.4$-K phase and a sharper ZBCP in the $3$-K
phase.  The latter was attributed to a nonchiral superconductivity
developing at the grain boundaries \cite{Sigrist01}.  An experimental
curve from Ref.\ \cite{Mao01} representing the $1.4$-K phase is shown
in Fig.\ \ref{fig:exper}.  It has the characteristic triangular shape
with a cusp at zero bias and agrees very well with our theoretical
curve calculated for the pairing potential (\ref{D5}).  To making the
comparison, we treated the horizontal and vertical scales and the
vertical offset (the background) of the theoretical curve as fitting
parameters.  The best fit corresponds to the gap $\Delta_0=0.8$ meV.

The good agreement between the theoretical and experimental curves is
a strong indication in favor of the $p$- or $f$-wave pairing potential
with horizontal lines of nodes in the 1.4-K phase.  Further
measurements showing the absence or presence of a ZBCP in the $c$-axis
tunneling could discriminate between the $p$- and $f$-wave cases.
Experiment \cite{Mao01} found that an additional sharp ZBCP with the
width of the order of 0.1 meV develops on top of the cusp at lower
temperatures 0.5 and 0.32 K.  It may be attributed to a contribution
from the $c$-axis tunneling, if the pairing potential is the $f$-wave
(\ref{D6}).

The described above scenario corresponds to the horizontal nodes in
the main electron band $\gamma$ of $\rm Sr_2RuO_4$.  In the
alternative scenario proposed in Ref.\ \cite{Zhitomirsky}, the main
band $\gamma$ is nodeless, but the two other bands $\alpha$ and
$\beta$ have horizontal lines of nodes.  In this case, one would
expect a superposition of a parabolic curve for the nodeless band and
triangular curves for the bands with nodes.  Perhaps the sharp ZBCP
developing at low temperatures 0.5 and 0.32 K could be interpreted as
the triangular curve corresponding to the gap of the order of 0.1 meV
in the $\alpha$ and $\beta$ bands.

\section{Conclusions}

We derived the analytical formula (\ref{tuncond}) for subgap tunneling
at low transparency of the barrier, which takes into account only the
contribution of the surface Andreev states.  This formula produces
tunneling curves in a simple and physically transparent way.  For the
chiral pairing potentials with vertical lines of nodes, tunneling
curves show double peaks, which originate from the turning points in
the energy dispersion of the surface Andreev states.  On the other
hand, for the chiral pairing potentials with horizontal lines of
nodes, we find a single triangular-shaped peak with a cusp at zero
bias, which results from the pile-up of the surface Andreev state at
zero energy caused by vanishing gap at the nodes.

Double peaks in tunneling conductance were observed experimentally in
some point contacts with $\rm Sr_2RuO_4$ in Ref.\ \cite{Laube00}.
However, such double-peaked spectra were associated with contacts with
high transparency of the barrier \cite{Laube00}, a regime that is not
addressed in our paper. On the other hand, no double peaks were
observed in experiment \cite{Mao01}, which only found a single
triangular-shaped peak at zero bias.  As shown in Fig.\
\ref{fig:exper}, this peak is very well fitted by our calculations for
horizontal lines of nodes.  It is also visually similar to the single
peak found for contacts with low transparency in Fig.\ 3 of experiment
\cite{Laube00} (which shows resistance, rather than
conductance). Thus, we conclude that the superconducting pairing
potential in $\rm Sr_2RuO_4$ most likely has horizontal lines of
nodes.

\emph{Note added in proof.}  Quantitative comparison shows that our
calculation for horizontal lines of nodes also fits the experimental
data of Ref.\ 29 very well.  The curves are shown at \\
http://www2.physics.umd.edu/\~{}yakovenk/talks/Sr2RuO4/

We are grateful to Ying Liu and Gernot Goll for sending us the
experimental data of Refs.\ 30 and 29, and to Yoshiteru Maeno, Ying
Liu, and Igor \u{Z}uti\'c for useful discussions.  This work was
supported by the Packard Foundation and by the NSF Grant DMR-9815094.


\end{document}